%% file: manuscript_main.tex
\documentclass[useAMS,usenatbib]{mn2e}
\newcommand{\electrons}{e$^{-1}$}
\newcommand{\teff}{$T_{\rm eff}$}
\newcommand{\rearth}{$R_{\oplus}$}
\newcommand{\rsun}{$R_{\odot}$}
\newcommand{\mearth}{$M_{\oplus}$}

\newcommand{\rstar}{$R_*$}

\newcommand{\delchi}{$\Delta \chi^2$}

\usepackage{aas_macros,amsmath,graphicx,longtable}

\title[SEAWOLF Search for Neptunes around Late-Type Dwarfs]{Trawling
  for transits in a sea of noise: A Search for Exoplanets by Analysis of
  WASP Optical Lightcurves and Follow-up (SEAWOLF)}

\author[Gaidos et al.]{E.~Gaidos,$^{1}$\thanks{E-mail:
    gaidos@hawaii.edu (EG)} D.~R.~Anderson,$^{2}$ S.~L\'{e}pine,$^{3}$ K.~D.~Col\'{o}n,$^{1}$ G.~Maravelias,$^{4}$ N.~Narita,$^{5}$ \newauthor
  E.~Chang,$^{1}$ J.~Beyer,$^{1}$  A.~Fukui,$^{6}$ J.~D.~Armstrong,$^{7}$ A.~Zezas,$^{4}$ B.~J. Fulton,$^{7,8}$ \newauthor A.~W.~Mann,$^{7}$ R.~G.~West,$^{9}$ and F. Faedi$^{9}$\\
  $^{1}$Department of Geology \& Geophysics, University of Hawaii at M\={a}noa, Honolulu, Hawaii 96822 USA\\
  $^{2}$Astrophysics Group, Keele University, Staffordshire ST5 5BG UK\\
  $^{3}$Department of Astrophysics, American Museum of Natural History, New York, NY 10024 USA\\
  $^{4}$Physics Department \& institute of Theoretical \& Computational Physics, University of Crete, 71003 Heraklion, Crete, Greece\\
  $^{5}$National Astronomical Observatory of Japan, 2-21-1 Osawa, Mitaka, Tokyo 181-8588 Japan\\
  $^{6}$Okayama Astrophysical Observatory, National Astronomical Observatory of Japan, Asakuchi, Okayama 719-0232, Japan\\
  $^{7}$Institute for Astronomy, University of Hawaii at M\={a}noa, Honolulu, Hawaii 96822 USA\\
  $^{8}$Las Cumbres Observatory Global Telescope Network, Goleta, CA 93117 USA\\
  $^{9}$Department of Physics, University of Warwick, Coventry CV4 7AL, UK}

\begin{document}

\date{Accepted 23 October 2013}

\pagerange{\pageref{firstpage}--\pageref{lastpage}} \pubyear{2013}

\maketitle

\label{firstpage}

\begin{abstract}
  Studies of transiting Neptune-size planets orbiting close to nearby
  bright stars can inform theories of planet formation because mass
  and radius and therefore mean density can be accurately estimated
  and compared with interior models.  The distribution of such planets
  with stellar mass and orbital period relative to their Jovian-mass
  counterparts can test scenarios of orbital migration, and whether
  ``hot'' (period $< 10$~d) Neptunes evolved from ``hot'' Jupiters as
  a result of mass loss.  We searched 1763 late K and early M dwarf
  stars for transiting Neptunes by analyzing photometry from the Wide
  Angle Search for Planets and obtaining high-precision ($\le
  10^{-3}$) follow-up photometry of stars with candidate transit
  signals.  One star in our sample (GJ~436) hosts a previously
  reported hot Neptune.  We identified 92 candidate signals among 80
  other stars and carried out 148 observations of predicted candidate
  transits with 1--2~m telescopes.  Data on 70 WASP signals rules out
  transits for 39 of them; 28 other signals are ambiguous and/or
  require more data.  Three systems have transit-like events in
  follow-up photometry and we plan additional follow-up observations.
  On the basis of no confirmed detections in our survey, we place an
  upper limit of 10.2\% on the occurrence of hot Neptunes around late
  K and early M dwarfs (95\% confidence).  A single confirmed
  detection would translate to an occurrence of $5.3 \pm 4.4$\%.  The
  latter figure is similar to that from Doppler surveys, suggesting
  that GJ~436b may be the only transiting hot Neptune in our sample.
  Our analysis of \emph{Kepler} data for similar but more distant
  late-type dwarfs yields an occurrence of $0.32\pm0.21$\%.  Depending
  on which occurrence is applicable, we estimate that the Next
  Generation Transit Survey will discover either $\sim$60 or
  $\sim$1000 hot Neptunes around late K and early M-type dwarfs.
\end{abstract}

\begin{keywords}
exoplanets --- planet formation --- transiting planets.
\end{keywords}

\section{Introduction \label{sec.intro}}

Not all exoplanets are detected equally.  A planet that transits its
host star has greater scientific value because its radius can be
determined and, because the orbital inclination is known, the
geometric ambiguity in Doppler estimation of the planet mass is
removed.  Spectroscopy of the star during a transit can reveal
absorption or scattering by the planet's atmosphere, if it has one.
The planet can also be occulted by the star, permitting differential
measurement of the planet's reflected or emitted flux.  These
observations can determine the planet's albedo and/or constrain the
efficiency with which heat is carried around the planet by rotation or
atmospheric circulation.

The most productive tool for detecting transiting planets has been the
\emph{Kepler} space telescope \citep{Borucki2010Sci}, data from which
has yielded more than 2000 confirmed or candidate discoveries.
However most of the systems discovered by \emph{Kepler}, as well as
those of the COnvection ROtation et Transits plan\'{e}taires (CoRoT)
satellite \citep{Carone2012} are faint ($V \sim 15$), making follow-up
observations difficult.  Many of these host stars are at kpc distances
and well above the Galactic plane, and may belong to an older, more
metal-poor population distinct from the Solar neighborhood and perhaps
hosting a different distribution of planets.

Ground-based surveys such as the Wide Angle Search for Planets
\citep[WASP,][]{Pollacco2006} and the Hungarian Automated Telescope
Network \citep[HatNET][]{Bakos2011} have discovered numerous giant
planets transiting brighter, nearby stars.\footnote{Transiting planets
  have also been identified by screening planetary systems detected by
  the Doppler method.  The first example (HD~209458) was found this
  way \citep{Charbonneau2000,Henry2000}, but this approach is limited
  by the pace of Doppler surveys and the small geometric probability
  that a planet will transit.}  Transiting geometries are uncommon and
such surveys must monitor many stars over large portions of the sky.
Because of the trade-off between field of view and telescope aperture,
these surveys are limited to the brightest stars and, due to Malmquist
bias, biased towards more luminous ones.  The sensitivity of such
surveys to smaller (non gas-giant) planets is limited by correlated
photometric error or ``red'' noise which does not decrease with the
square root of the number of observations \citep{Pont2006,Smith2007}.
Earth's rotation means that surveys performed from a single site have
restricted observing windows and are only efficient at detecting
planets on short-period orbits ($\le 10$~d).  For these reasons,
ground-based surveys have been most successful at detecting giant
planets on close orbits around F and G stars.\footnote{Wide-field
  surveys must also contend with a high false positive rate by blends
  of bright stars with fainter eclipsing binaries.}

M dwarf stars have less than half the radius of their solar-type
cousins, permitting the detection of concomittantly smaller planets
for a given photometric sensitivity.  Although such stars tend to be
fainter and observed with poorer photometric precision, the net
balance of these two effects can still favor cooler stars: this
calculus motivates the MEarth transit survey for planets as small as
Earth around late M-type dwarfs \citep{Charbonneau2009a,Berta2012}.

K- and early M-type dwarfs represent an intermediate region of
discovery space for transiting planet surveys.  While ground-based
detection of Earth-size planets around such stars is not feasible, it
is possible to detect Neptune- or even super-Earth-size companions, at
least on close-in orbits.  Indeed, HAT-P-11b (4.3\rearth{}, $P =
4.89$~d) transits a K4 dwarf, and HAT-P-26b (6.3\rearth{}, $P =
4.23$~d) orbits a K1 dwarf \citep{Hartman2011}.\footnote{Two other
  Neptune-size planets, both around early-type M dwarfs, were detected
  first by Doppler, then later found to transit: GJ~436b
  \citep{Gillon2007}, and GJ~3740b \citep{Bonfils2012}.}

The occurrence and properties of short-period or ``hot'' Neptunes are
of considerable theoretical interest.  Attempts to explain an apparent
correlation between the occurrence of giant planets and stellar mass
also predict an inverse relation with elevated numbers of Neptunes
(i.e., ``failed Jupiters'') around low-mass stars
\citep{Laughlin2004}.  Hot Neptunes could form by accretion of
rocky/icy planetesimals beyond the snowline and subsequent migration
to the inner edge of the protoplanetary disk \citep{Mordasini2009}.
However, \citet{McNeil2010} find that this scenario cannot explain the
observed size distribution of close-in planets.  Alternatively,
planetesimals or protoplanets could migrate first, followed by
accretion in place \citep{Brunini2005,Hansen2012}.  Finally,
evaporation of mass from close-in giant planets has been proposed as
an alternative formation mechanism for hot Neptunes
\citep{Baraffe2005,Boue2012}.  These three different pathways predict
objects that are enriched in ice, rock, and gas, respectively.  Hot
Neptunes may be especially useful to test models of planet formation
because both mass and radius can be accurately measured (by Doppler
and transit, respectively) and these parameters are informative about
the relative amounts of rock, ice and gas in the planet
\citep{Rogers2011}.  In contrast, the masses of Earth-size planets are
too small to accurately measure and the radii of Jupiter-size planets
are insensitive to mass due to support by electron degeneracy
pressure.

We used data from the WASP survey to search for short-period Neptunes
around a sample of low-mass stars (the SEAWOLF survey).  Because this
search pushes the envelope of WASP performance, we adopted a
multistage search strategy:
\begin{itemize}
\item We selected late K- and early M-type dwarf stars observed by the
  WASP survey; in principle, smaller planets should be detectable
  around these smaller stars.
\item We identified candidate transit signals, relaxing the
  signal-to-noise criterion for initial selection.  This potentially
  includes smaller transit signals, but also large numbers of false
  positives.
\item We predicted candidate transits using the WASP-generated
  ephemerides and screened these with precision photometry obtained at
  1--2~m telescopes.
\end{itemize}
We describe the WASP data and our follow-up observations and reduction
in Section \ref{sec.methods}, and our catalog of candidate transiting
systems and the results of the follow-up program in Section
\ref{sec.results}.  We place limits on the occurrence of hot Neptunes
around stars in our sample in Section \ref{sec.analysis}, and discuss
the implications for theory as well as prospects for future transiting
planet surveys in Section \ref{sec.discussion}.

\section[]{Observations and Methods} \label{sec.methods}

\subsection{Sample construction and stellar parameters} \label{subsec.sample}

For our search sample we identified late-type (K4 to M4) dwarf stars
in the inaugural (2004) fields of the WASP-North survey
\citep{Christian2006}.  We chose stars from the SUPERBLINK proper
motion catalog \citep{Lepine2005} with optical-to-infrared colors $V-J
> 2$ consistent with late K- and M-type stars, and reduced proper
motions $H_J \equiv J + 5 \log \mu + 5$ (a proxy for absolute
magnitude) that place them on the dwarf color-magnitude locus, thus
excluding K and M giants \citep{Lepine2011}.  We restricted the sample
to $V < 14$ because at fainter magnitudes the number of background
stars with $\Delta m < 5$ falling within the same WASP photometric
aperture significantly exceeds one.  Such stars could produce false
positives if they are eclipsing binaries.  We also imposed a $J < 10$
cut to retain those stars for which high-precision, high cadence (few
minute) photometry in the near-infrared ($z$ or $JHK$ passbands) could
be performed on 1--2~m telescopes.  Based on parallaxes (astrometric
wherever available, photometric otherwise), the most distant stars in
our survey are at $\approx$100~pc.  The closest star is Laland 21185,
only 2.5~pc away.

To estimate the properties of these stars, we adopted the empirical
relations between $V-J$ color, effective temperature \teff{}, stellar
radius $R_*$, and stellar mass $M_*$ for solar-metallicity K and M
stars in \citet{Boyajian2012ApJ757}\footnote{We used 2MASS $J$
  magnitudes while \citet{Boyajian2012ApJ757} used Johnson $J$
  magnitudes.  However, the CIT photometric system is closely related
  to the Johnson system and $J_{\rm CIT} \approx J_{\rm 2MASS} -
  0.065\left(J-K\right)_{\rm 2MASS} + 0.038$ \citep{Skrutskie2006}.
  Since late K-early M dwarfs have $J-K \approx 0.8$, the difference
  in $V-J$ color is only 0.014 magnitudes and was ignored.}.
According to these relations, $V-J = 2$ corresponds to $R_* \approx
0.71R_{\odot}$, \teff{} $\approx$ 4550~K, and a spectral subtype of K4
\citep{Cox2000}.  The coolest stars in our sample have $V-J \ge 4.5$
and should have M4 spectral types, with \rstar{} $\approx$ 0.25\rsun{}
and \teff{} $\approx$ 3300~K.  The reddest star ($V-J$ = 5.39) is the
M4.5 dwarf GJ~3839.

\subsection{WASP Observations and Sources} \label{subsec.wasp}

\begin{figure}
\includegraphics[width=84mm]{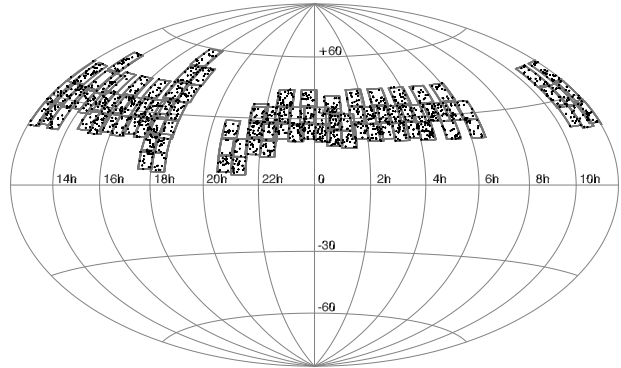}
\caption{Locations of the 2004 inaugural fields of the WASP-North
  survey and our selected SUPERBLINK K and M dwarf stars.}
 \label{fig.fields}
\end{figure}

\begin{figure}
\includegraphics[width=84mm]{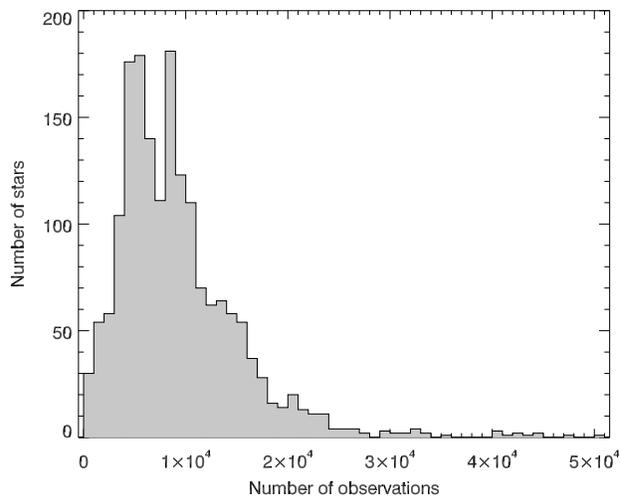}
\caption{Distributions of number of WASP observations per star in our
  sample.  The median number of observations per star is
  8160. \label{fig.nobs}}
\end{figure}

We identified 1849 SUPERBLINK stars satisfying our criteria in the
inaugural (2004) fields of the WASP-North survey.  These 102 fields
cover 4750 sq. deg. at declinations between +4.9 and +59.3
deg. (Fig. \ref{fig.fields}).  Stars were matched with sources
generated by photometering WASP images with a circular aperture of
radius 3.5~pixels (48 arc~sec).  1763 WASP sources were matched to our
selected SUPERBLINK stars; the median angular separation is 0.22
arc~sec and the 95th percentile separation is 3.4 arc~sec.  Forty-six
WASP sources were each matched to two SUPERBLINK stars.  Of the 1763
matched sources, 1743 have more than 500 data points (the minimum
required for lightcurve analysis) and the median number of
observations is 8160 (Fig. \ref{fig.nobs}).

\subsection{Light Curve Analysis} \label{subsec.lightcurves}

WASP lightcurves were processed to correct for systematic errors
\citep{Tamuz2005} and remove trends \citep{Kovacs2005}.  The latter
step eliminates many artifacts with periods equal to rational
multiples of 1~d.  The light curves were then analyzed with the HUNTER
hybrid search algorithm that incorporates the box-least squared
algorithm \citep{Kovacs2002} and which is described in
\citet{CollierCameron2006}.  HUNTER searched for transit-like signals
with periods of 0.3--30~d.  Four criteria were applied to these
signals: (i) mean flux $>3$ microVegas ($m > 13.8$); (ii) periods not
within 5\% of 1 or 0.5~d (see Section \ref{subsec.selection}); (iii)
signal detection efficiency $>6$ \citep{Kovacs2002}; (iv) at least
three candidate transits.

Up to five periodic signals were identified for each source satisfying
these criteria, and a total of 4364 signals were identified among 1130
stars.  HUNTER calculated the signal-to-red noise (SRN) and \delchi{}
parameter for each signal, where the latter is the decrease in
$\chi^2$ provided by the best-fit transit model relative to a constant
light curve.  In the case of pure white noise \delchi{} is the square
of the signal-to-white noise ratio \citep{CollierCameron2006}.  Thus
SRN and \delchi{} allow us to select based on the significance of a
signal with respect to both the red noise and white noise properties
of the data.

\subsection{Selection of Candidate Transiting Systems} \label{subsec.selection}

\begin{figure}
\includegraphics[width=84mm]{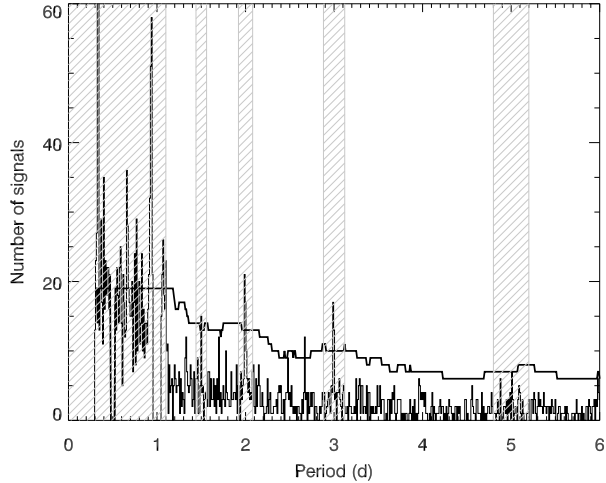}
\caption{Period distribution of signals from the WASP HUNTER pipeline.
  The upper solid curve is a significance threshold ($p = 10^{-4}$)
  based on the Poisson statistics of a running mean ($n = 50$).
  Clusters of artifacts are present at rational multiples of 1~d.  The
  hatched regions indicate exclusion zones around these periods and at
  $<$1.1~d; signals within these zones were rejected.}
\label{fig.periodhist}
\end{figure}

We next applied cuts with period, SRN, \delchi{} and ellipsoidal
signal-to-noise ratio (a measure of the continuous variation of the
signal over the period) to the 4364 HUNTER-identified signals to
screen artifacts and astrophysical false positives (i.e., close
binaries).  Because of Earth's rotation, observations from a single
longitude like those of WASP-North can contain artifacts with periods
near 1~d and integer ratios thereof.  Furthermore, aliasing with the
lunar cycle (29.5~d) produces a dispersion of a few percent around
each rational period.  Based on the distribution of signals (mostly
artifacts) generated when no detrending is performed (see above), we
removed signals with periods below 1.1~d and within 5\% of 3/2, 2, 3,
and 5~d (Fig. \ref{fig.periodhist}).  (A peak at 4~d is not
statistically significant.)  There is also a peak in the period
distribution of signals at 8/3~d.  This peak appears significant and
is apparently one of a series of undertones (multiples) of the strong
artifact at 1/3~d, the harmonic closest to the duration of a summer
night at the WASP-North site (and which is removed along with all
other signals below 1.1~d).  However, we did not a priori remove the
1/3~d peak.

\begin{figure}
\includegraphics[width=84mm]{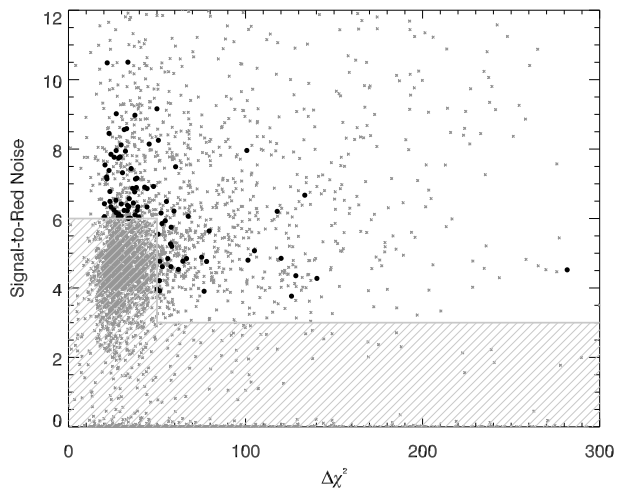}
\caption{Signals from the WASP HUNTER pipeline; signal-to-red noise
  ratio vs. $\Delta \chi^2$.  Only signals outside the hatched zone
  were considered.  The final candidate transiting systems are
  indicated by the large black points.} \label{fig.snrvschi2}
\end{figure}

The distribution of signals with SRN and \delchi{} is strongly
concentrated at SRN $\sim$4.5 and \delchi{} $\sim$30
(Fig. \ref{fig.snrvschi2}).  We assumed that these are nearly all
artifacts or astrophysical false positives and that the clustering is
a result of the selection criteria applied in Section
\ref{subsec.lightcurves}.  We retained signals with $(SRN > 6) \cup
\left(SRN > 3 \cap \Delta \chi^2 > 50 \right)$ (outside the hatched
zone of Fig. \ref{fig.snrvschi2}).  We also excluded signals with an
ellipsoidal SNR $>8$: these are possible close binaries
\citep{CollierCameron2006}.  The remaining 901 signals were further
screened with the following criteria: (i) observations had to
completely span, ingress to egress, at least three putative transits;
(ii) the putative transit duration $\tau$ had to be within a factor of
two of the value for a planet on a circular orbit with zero impact
parameter around a star with radius 0.6$R_{\odot}$ (typical of a late
K dwarf), i.e. $\tau = {\rm 1~hr~}(P/{\rm 1~d})^{1/3}$, where $P$ is
the Keplerian period; and (iii) the putative signal could not
obviously be an artifact produced by periodic gaps or changes in noise
level in the data.  This left 92 candidate signals from 80 stars.

\subsection{Follow-up Photometry} \label{subsec.followup}

We used the ephemerides generated by the HUNTER pipeline to predict
transits for our 92 candidates.  The precision of the predicted
transit center depended mostly on the precision of the period
determination, and was generally $\pm$1~hr.  Follow-up photometry of
some of these candidate transit events was obtained with 1--2~m
ground-based telescopes.  Details of the telescopes are reported in
Table \ref{tab.telescopes} and of the observations in Table
\ref{tab.observations}.  In general, we selected events to observe if
the predicted transit center occurred when the star was at an airmass
below 1.7, and at least 3~hr from sunset or sunrise.  Ideally, we
observed the entire transit window ($\pm 1\sigma$) as well as an hour
before and after ingress/egress.  However in many cases this was not
possible.  The minimum detectable transit depth $\delta_d$ and
completeness $C$ of these observations are calculated in Section
\ref{sec.analysis}.

Although the details of the observing strategy and data reduction
varied with telescope and instrument, there were several
commonalities:

{\it Defocused imaging photometry:} A telescope was grossly defocused
to produce a ``doughnut''-shaped point spread function (PSF) several
tens of pixels in diameter.  Such ``doughnuts'' are out-of-focus
images of the primary mirror.  Defocusing permitted a signal $\gg 1
\times 10^6$ \electrons{} to be acquired in each integration, reducing
Poisson error to $< 10^{-3}$.  It also minimized error from image
motion or changes in the distribution of the signal convolved with
detector flat-fielding errors \citep[e.g.][]{Southworth2009,Mann2011}.
Circular aperture photometry was performed on the defocused images of
the target and several comparison stars in multiple iterations.  In
each iteration the centroid of the stellar image within the aperture
was computed and used as the aperture center.

{\it Optimized pointings:} The signal from a star of interest must be
divided by that from one or more comparison stars to remove variations
in atmospheric transmission.  The number and relative brightness of
comparison stars limits the precision of ground-based photometry.  We
chose pointings that maximized the number of comparison stars similar
in brightness to the target star.  We also avoided rings in the flat
field due to dust particles near the focal plane.  These can change
between nights or even during observations, introducing flat-field
error.  

{\it Comparison star selection:} Each comparison star was compared
with all the others to identify and exclude variables.  A comparison
signal was calculated from the weighted sum of the remaining reference
stars, where the weights were chosen to minimize the RMS of the
normalized target light curve outside the predicted transit window.

{\it Lightcurve detrending:} We performed linear regressions of each
normalized light curve with airmass, position of the centroid, and
variance in the distribution of the target star signal over the
point-spread function.  The first was to remove second-order
extinction effects due to differences in the spectra of target and
reference stars \citep{Mann2011}.  The second partly removes
flat-field errors introduced when the defocused images move due to
imperfect guiding or absence of guiding.  The third compensates for
any non-linearity in the response of the detector which would scale
with the variance in the light distribution.

\section[]{Results} \label{sec.results}

\subsection{Candidate Signals} \label{subsec.candidates}

The final catalog of 92 candidate signals from 80 stars is presented
in Table \ref{tab.candidates}.  The ``A'', ``B'', or ``C''
designations indicate different signals from the same star.  Based on
the depths of the putative WASP transits and the estimated radii of
these stars (see Section \ref{subsec.sample}) the median transiting
planet radius would be $\sim$4 \rearth{}, i.e. Neptune-size.  However,
we expect that the large majority of these signals are artifacts or
astrophysical false positives and not transiting planet.  In Table
\ref{tab.candidates} we report the status of each candidate based on
the follow-up photometry acquired to date: N = none, ? = ambiguous or
requires further observations, X = eliminated, A = candidate
transiting system.  We have follow-up observations of 70 signals and
we ruled out 39 signals and designated 28 as ambiguous or lacking
sufficient data.  In general, systems where the completeness $C$ of
our follow-up observations is $>$80\% (as calculated in Section
\ref{subsec.completeness}), and no transit-like event was observed,
were ruled out, and systems with one or more observations but where
completeness was $<$80\% were designated as ``?''.  There are five
exceptions to the rule: Two systems (03571+3023 and 14162+3234) have
$C \approx 0.77$ but were ruled out.  Six systems (15015+2400,
16389+3643B, 21302+2312A, 21409+1824, 22085+1425A and 22085+1425B)
have $C > 0.8$ but the predicted event was close to the beginning or
end of an observing window, a possible event was observed
significantly before the predicted time or different observations had
conflicting results: these are designated as ``?''.

\subsection{Transit Candidates} \label{subsec.transitcand}

\begin{figure}
\includegraphics[width=84mm]{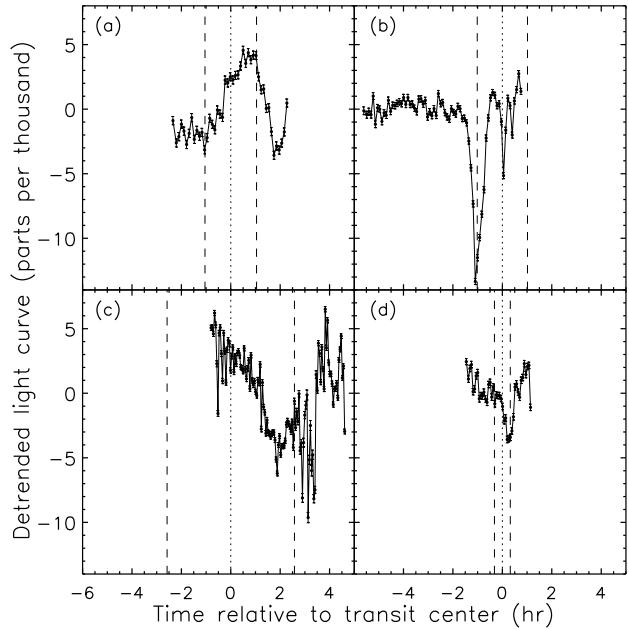}
\caption{Detrended lightcurves from follow-up observations of four
  stars containing a transit-like event.  The error bars show the
  $1\sigma$ errors from Poisson noise only.  The vertical dotted lines
  mark the predicted transit time and the vertical dashed lines mark
  $\pm$ one standard deviation.  The stars and UT epochs are (a)
  03571+3023 on 16 Sept 2012, (b) 16442+3455 on 3 May 2013, (c)
  17378+2257 on 24 April 2013, and (d) 18075+4402 on 27 April 2013.}
 \label{fig.lightcurves}
\end{figure}

Follow-up observations of four signals produced light curves that
contain a transit-like signal: 03571+3023, 16442+3455, 17378+2257, and
18075+4402 (Fig. \ref{fig.lightcurves}).  We have continued to observe
predicted events for these stars to verify or rule out possible
transits.  03571+3023 is variable lightcurves are not consistent
between predicted events and one is flat, causing us to rule out this
system.  16442+3455 (Ross 813) was mis-assigned to a white dwarf in
the catalog of \citet{McCook1987} but its colors and luminosity are
clearly those of a late K or early M dwarf.  17378+2257 (GJ 686.1AB,
HIP 86282) consists of a pair of dwarfs that have $V-J \approx 3$ and are
designated as M0 stars in the Gliese catalog but listed as K5 in
\citet{Reid1995}.  The molecular indices reported by \citet{Reid1995}
and a spectrum obtained by us with the Mark III spectrograph on the
MDM 1.3~m McGraw-Hill telescope suggest a spectral type between K7 and
M0.  These stars are an X-ray source \citep{Hunsch1999} and the
S-index values of the Ca II HK lines in their spectra
\citep{Duncan1991} suggest the stars are comparatively active
\citep{Isaacson2010}, but H$\alpha$ is not observed in emission
\citep{Young1989}.

\subsection{GJ~436b} \label{subsec.gj436b}

\begin{figure}
\includegraphics[width=84mm]{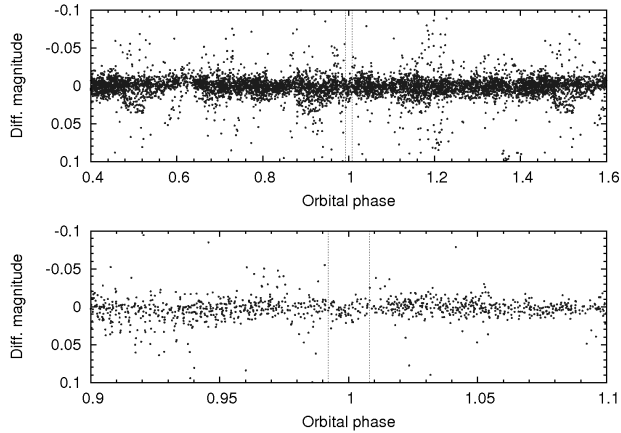}
\caption{WASP lightcurve of GJ~436, which hosts a hot Neptune on a
  2.64~d orbit.  The data has been phased according to the established
  ephemeris of the planet and the transit is marked by the vertical
  lines.  The bottom panel plots on an expanded
  scale. \label{fig.gj436}}
\end{figure}

GJ~436 aka SUPERBLINK star PM~I11421+2642 or WASP source
J114210.54+264230.4 is in our sample.  The transit signal from its
4.3\rearth{} planet \citep{Gillon2007} was not detected by the WASP
pipeline and in fact no candidate signals were identified from this
star.  One possible explanation for the system's omission is that only
one season of WASP data was obtained and the star fell 2.4~deg from
the center of the field of view and thus was vignetted.  Another
contributing factor is the planet's high transit impact parameter ($b
= 0.85$), which makes the transit unusually short.  The data is also
exceptionally noisy: the 5$\sigma$-filtered RMS is 1.2\%, consistent
with the nominal photometric error of 1.4\% and about 2.5 times the
typical value for a $V = 10.7$ star (see Eqn. \ref{eqn.rednoise}
below).  The transit is only marginally apparent even after the data
is correctly phase-folded (Fig. \ref{fig.gj436}).  Although transits
of GJ~436b were not detected by WASP, the inclusion of this system in
our sample raises the question of whether we should expect additional
hot Neptunes or whether this is the only such transiting system.  For
these reasons we carry out our statistical analysis for zero and one
detections (Sec. \ref{sec.analysis}).

\section[]{Analysis} \label{sec.analysis}

\subsection{Estimation of WASP detection limits}

To place statistical constraints on the occurrence of hot Neptunes we
calculated (i) the ability of HUNTER to detect planets in WASP
lightcurves as a function of planet radius and orbital period, and
(ii) the completeness with which our follow-up observations can rule
out candidate transit signals (Section \ref{subsec.completeness}).
Our criteria for WASP/HUNTER detection is the same as that applied to
the data: signal-to-red noise SRN $> 6$, or SRN $> 3$ and
signal-to-\emph{white} noise $> \sqrt{50}$.

The transit signal $\delta = (R_p/R_*)^2$, where $R_p$ is the radius
of the planet.  The red-noise error in the mean of $N$ observations in
the transit interval is $\sigma_1 N^{-\gamma}$, where $\sigma_1$ is
the error in a single WASP measurement of a given star and the index
$\gamma \approx 0.5 - 0.05 \times (15-V)$ \citep[based on Fig. 2
in][]{CollierCameron2006}.  The white-noise error is taken to be
$\sigma_1 N^{-0.5}$.  We constructed an empirical formula for
$\sigma_1$ based on Fig. 2 in \citet{CollierCameron2006}:
\begin{equation}
\label{eqn.rednoise}
\sigma_1 = 2.5 \times 10^{-3} \sqrt{1 + 8 \times 10^{\left(V-13\right)/5}}.
\end{equation}
Assuming near-circular orbits for these close-in planets, the mean
number of observations falling within a transit is taken to be $N =
n\tau/P$, where $\tau$ is the duration of the transit and $n$ is the
total number of observations.  The transit duration in hours is
\begin{equation}
\tau \approx 0.075R_*P^{1/3}\sqrt{1-b^2},
\end{equation}
where $R_*$ is in solar units, $P$ is in days, and $b$ is the impact
parameter (taken to be zero here).

\begin{figure}
\includegraphics[width=84mm]{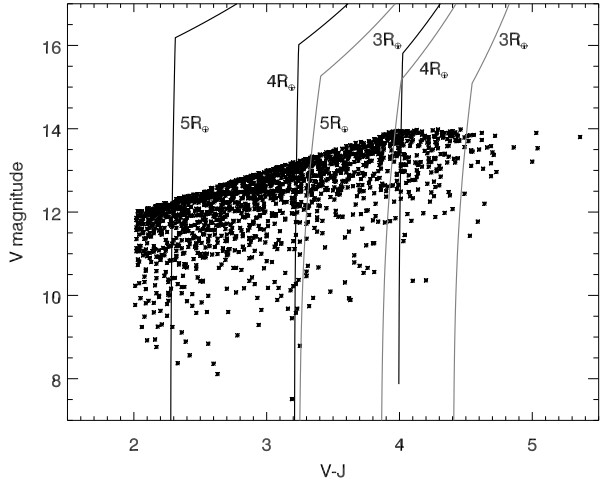}
\caption{Expected WASP detection limits for Neptune-size planets
  around late K and M dwarf stars, plotted vs. $V-J$ color (a proxy
  for \teff{} and spectral type) and $V$ magnitude.  Transiting
  planets of specified radius (3, 4 or 5 \rearth{}), and orbital
  period (1.2~d, black curves, or 10~d, grey curves) should be
  detectable around stars to the right and below each curve.  The
  stars of the SEAWOLF survey are plotted.  Circular orbits, an impact
  parameter of zero and the median number of observations in our
  survey sample (8160) are assumed for these calculations.}
 \label{fig.limits}
\end{figure}

Figure \ref{fig.limits} plots the limiting $V$ magnitude for detecting
a transiting Neptune-size (3, 4, or 5\rearth{}) planet around a dwarf
star in the WASP survey as a function of stellar $V-J$ color, our
proxy for \teff{} and stellar radius on the main sequence, for $P
=$1.2~d or 10~d (see Section \ref{sec.occurrence} for a justification
for this range).  The $V-J$ and $V$ of stars in our survey catalog are
overplotted.  Figure \ref{fig.limits} shows that WASP should be able
to detect planets somewhat larger than Neptune (5--6\rearth{}) around
nearly all of the stars in our sample, and planets slightly smaller
than Neptune ($\sim$3\rearth) around the coolest (M dwarf) stars, but
only if they orbit quite close to their host.  For orbital periods of
10~d only the largest Neptunes will be detectable around the M dwarfs.

The break in the slope of the detection contours at $V \approx 15$--16
in Fig. \ref{fig.limits} is a result of a transition from a photon- or
counting noise-limited regime to a red noise-dominated regime. Among
stars with a fixed radius ($V-J$ color) and $V > 15$--16, detection
improves with brightness.  However, for $V < 15$--16 correlated noise
becomes important (decreasing $\gamma$ with brighter $V$).  For a
fixed $N$, this means that detection requires a lower $\sigma_1$ and
hence an even brighter $V$.  This positive feedback means that for a
given stellar radius, planets below a certain size cannot be detected,
regardless of apparent magnitude.  This is a widely-appreciated
limitation of ground-based surveys \citep{Pont2006,Smith2007}.

\subsection{Estimation of follow-up completeness} \label{subsec.completeness}

To evaluate the significance of non-detection or detection of transits
in our follow-up observations, we calculated the completeness,
i.e. the probability that a transit with the characteristics of the
WASP candidate would be detected, and the false-alarm probability,
i.e. the probability that such an event would be erroneously
identified in our data in the absence of an actual transit.

Each follow-up observation of a candidate transit event yielded a
normalized, de-trended lightcurve, plus errors based purely on
counting statistics (Poisson or photon noise).  False alarm
probability and completeness were calculated by constructing two sets
of Monte Carlo realizations of the data, the first set with no transit
signal added, and the second containing a transit signal equal in
depth to the WASP candidate.  The first set was used to set the
detection threshold, i.e. the transit depth corresponding to a false
alarm probability (FAP) of 0.01.  This means there is a 1\%
probability that a signal exceeding this threshold would be
erroneously discovered in a lightcurve with these noise properties but
containing no transit.  We used the second set of Monte Carlo
realizations plus the detection threshold determined from the first
set to estimate the completeness or the recovery rate in follow-up
photometry of transit signals with the properties of the WASP
candidate.

To account for the effect of correlated or ``red'' noise on transit
detection, we computed the discrete autocorrelation function $A_k$ of
the actual data and used this to construct artificial light curves
$s_i$ of pure noise:
\begin{equation}
s_i = s \displaystyle\sum_j A_{i-j}w_j,
\end{equation}
where $w_i$ is a white noise pattern and $s$ is chosen so that $s_i$
has the same total noise RMS as the actual signal.

Our simple transit model used a linear limb darkening
law\footnote{More complex limb-darkening laws are widely used but a
  linear law is completely adequate for creating and ``detecting''
  fake transits at low signal-to-noise.} and the ``small planet''
approximation ($R_p \ll R_*$) such that the transit signal is
\begin{equation}
f(t) = \delta \left[1- u \left(1-\mu \right)\right],
\end{equation}
for $r < 1$, where $\mu = \sqrt{1-r^2}$, $u$ is the linear
limb-darkening coefficient, the dimensionless radial coordinate is
\begin{equation}
r =  \sqrt{\left(2(t-t_c)/\tau\right)^2 + b^2},
\end{equation}
and $t_c$ is the transit center time.  Based on the median estimated
\teff{} of our sample (4570~K) and assuming solar metallicity, we
adopted values of $u=$0.80, 0.72, and 0.51 for Johnson $V$ and $R$ and
Tiede $J$ bandpasses, respectively \citep{Claret2000}, and $u = $0.75,
0.65, and 0.58 for Sloan \emph{riz} bandpasses, respectively
\citep{Claret2004}.  To calculate the transit duration $\tau$ we
assumed a circular orbit and a stellar radius based on $V-J$ and
\citet{Boyajian2012ApJ757} (see Section \ref{subsec.sample}).  For
each Monte Carlo realization, we drew a fixed value of impact
parameter $b$ from a uniform distribution limited to $\sqrt{3}/2$
(beyond which the transit duration is half the maximum value,
resulting in exclusion from our sample).

To generate a distribution of false-positive transits, we fit the
transit model to each transit-free light curve using the non-linear
least-squares routine MPFIT \citep{Markwardt2009}, with $t_c$ and
$\delta$ as free parameters.  For the fit, an initial value of
$\delta$ was chosen from a uniform distribution between 0 and 0.002.
An initial value of $t_c$ was chosen from a normal distribution with a
standard deviation equal to the transit prediction error, and limited
to the observation window.  Cases where the fitted depth was negative
or the transit was more than two standard deviations from the
predicted transit center were not counted, as these would have been
excluded from the actual survey.  We determined the 99 percentile
value of the transit depth, corresponding to a false alarm probability
of 0.01.  This is our adopted detection limit $\delta_d$.  This value
was converted to an equivalent planet radius using the stellar radius,
and we also computed a corresponding SNR detection threshold based on
the white-noise RMS of the light curve.

To calculate the completeness, we added artificial transits to the
noise-only light curves and attempted to recover them.  Each transit
was modeled as described above, using the WASP candidate transit
depth, a uniform distribution for $b$ between 0 and $\sqrt{3}/2$ and a
normal distribution for $t_c$.  We then repeated the fitting process
described above.  To initially ``detect'' the transit, we smoothed
each light curve with a boxcar filter having a width equal to the
expected transit duration.  The minimum of this lightcurve became the
initial guess at $t_c$ in a fit with MPFIT.  We calculated the
fraction of recovered transit depths that exceeded $\delta_d$,
rejecting fits with $t_c$ deviating from the actual value by more than
two standard deviations.  This fraction is our estimated completeness
$C$.  Table \ref{tab.observations} reports values of $\delta_d$ and
$C$ for each follow-up observation.

\subsection{Planet Occurrence} \label{sec.occurrence}

Our observations constrain the intrinsic occurrence $f$ (planets per
star, or in the limit of few planets, fraction of stars with planets)
of close-in Neptune- to Saturn-size planets around late K and early M
dwarf stars in the solar neighborhood.  A standard procedure to
estimate $f$ is to maximize a likelihood function that is the product
of the probabilities of detections and non-detections.  Our
multi-stage observational campaign required us to consider how we
defined detections and non-detections. Specifically, our sample
includes:
\begin{itemize}
\item Stars with no transit-like signal found in WASP data: These were
  counted as non-detections.
\item Stars with a transit-like signal identified in WASP data but
  which were not screened with follow-up observations: These were
  considered as unconfirmed detections.
\item Stars with transit-like signals in WASP data which our follow-up
  observations have ruled out as transit candidates with some
  completeness $C$: These are considered possible non-detections or
  unconfirmed detections.
\item Stars with WASP signals that our follow-up photometry indicate
  are viable transit candidates: Given sufficient follow-up, these
  could become confirmed detections.
\end{itemize}
Following \citet{Gaidos2013c} we generalized the likelihood formalism
as an empirical Bayes/marginalized likelihood analysis in which the
occurrence rate $f$ is a ``hyperparameter'' of the prior probability
that a star hosts a detectable transiting planet \footnote{Strictly
  speaking, the fraction of stars with such planets, which is equal to
  the number of such planets per star if the possibility of multiple
  planets in this restricted range of radii and orbital periods is
  neglected.}.  This prior is $\langle d_i\rangle f$, where $\langle
d_i \rangle$ is the probability that a planet transits and is detected
with the criteria in Section \ref{subsec.selection}, marginalized over
the distributions of planet radius and orbital period.  The
log-likelihood is
\begin{equation}
\begin{split}
\ln \mathcal{L} = \displaystyle \sum_i^{ND} \ln \left(1-f\langle d_i\rangle \right) +  \displaystyle \sum_k^{CD} \ln \left[(1-F_k)fd_k\right]\\ + \displaystyle \sum_j^{UD} \ln  \left[(1-C_j)fd_j + C_j\left(1-f\langle d_j \rangle\right)\right],
\end{split}
\end{equation}
where the summations are over non-detections (ND), confirmed
detections (CD), and uncomfirmed detections (UD), $C_j$ is the
completeness of the follow-up observations that do not find a transit,
and $F_k$ is the false-alarm probability for detections confirmed by
our follow-up observations.  In the case of multiple observations of
the same system we adopt the largest value of $C_j$.

If $fd \ll 1$ and $F \ll 1$, then
\begin{equation}
\begin{split}
\label{eqn.occurrence}
\ln \mathcal{L} \approx N_{CD} \ln f - f \displaystyle \sum_i^{ND}\langle d_i\rangle  \\ + \displaystyle \sum_j^{UD} \ln  \left[(1-C_j)fd_j + C_j\left(1-f\langle d_j \rangle\right)\right], 
\end{split}
\end{equation}
where $N_{CD}$ is the number of confirmed detections.  If $C_i$ is not
small for all stars (not the case here) then this can be further approximated as:
\begin{equation}
\begin{split}
\ln \mathcal{L} \approx N_{CD} \ln f + \displaystyle \sum_j^{UD} \ln C_j \\
- f \left[\displaystyle \sum_i^{ND}\langle d_i\rangle + \displaystyle \sum_j^{UD} \left(C_j\left<d_j\right>-\frac{1-C_j}{C_j}d_j\right)\right] .
\end{split}
\end{equation}
Only the first two terms depend on $f$, and from these one readily
derives the most likely value:
\begin{equation}
  f_* = N_{CD} \left[\displaystyle \sum_i^{ND}\langle d_i\rangle + \displaystyle \sum_j^{UD} \left(C_j\left<d_j\right>-\frac{1-C_j}{C_j}d_j\right)\right]^{-1}
\end{equation}
If no transits are confirmed the most likely value of $f$ is zero.

The detection probability $d_i$ for a given candidate transit signal
is the product of a geometric factor $d_i^{\rm geo}$ and a detection
probability $d_i^{\rm det}$.  Assuming circular orbits,
\begin{equation}
\label{eqn.geofactor}
d_i^{\rm geo} \approx 0.238 R_*M_*^{-1/3}P^{-2/3},
\end{equation}
where the stellar parameters are in solar units and $P$ is the signal
period in days.  $d_i^{\rm det}$ is estimated by computing both the
SRN and \delchi{} for the given $\delta$, $P$, and a
uniformly-distributed range of impact parameters, and determining the
fraction of these which satisfy our selection criteria (Section
\ref{subsec.selection}).  The SRN is given by $\delta N^{\gamma} /
\sigma_1$, $\Delta \chi^2 = \delta^2 N/\sigma_1^2$, and $\gamma$ and
$N$ are estimated as before.  All cases with $b > \sqrt{3}/2$ were
excluded because of the restriction on candidate transit duration
(Section \ref{subsec.selection}).

To calculate $\langle d \rangle$ we assumed a power-law distribution
over radius $R_{\rm min} < R_p < R_{\rm max}$ with index $\alpha$, and
a flat log distribution with orbital period $P_{\rm min} < P < P_{\rm
  max}$ \citep{Cumming2008,Howard2012}, i.e.
\begin{equation}
dN = \frac{f\left(R_p/R_{\rm min}\right)^{-\alpha}d\ln R_p\;d\ln P}{\alpha \ln \left(P_{\rm max}/P_{\rm min}\right)}
\end{equation}
where $R_p > R_{\rm min}$ and $P_{\rm min} < P < P_{\rm max}$.  We
calculated the minimum detectable planet radius $R_{\rm det}$ and the
fraction of planets that would be detected, i.e.
\[
d^{det} =
\begin{cases}
  \frac{R_{\rm det}^{-\alpha}-R_{\rm max}^{-\alpha}}{R_{\rm min}^{-\alpha}-R_{\rm max}^{-\alpha}}, & \text{if } R_{\rm min} < R_{\rm det} < R_{\rm max}\\
  1, & \text{if } R_{\rm det} < R_{\rm min} \\
  0, & \text{if } R_{\rm det} > R_{\rm max} \; \; \; \; \; \; \; \; \; \; \; \; \; \; (10b)
\end{cases}
\]
We marginalized over $P$ and $b$ assuming logarithmic and uniform
distributions, respectively, and excluding values of either parameter
that were also excluded during our selection of candidate transit
signals, i.e. $b > \sqrt{3}/2$ and $P < 1.1$~d or periods within 5\%
of artifacts (Section \ref{subsec.selection}).

We adopted $R_{\rm min} = 3$\rearth{} and $R_{\rm max} = 8$\rearth{},
i.e. slightly smaller than Neptune and Saturn, respectively.  We
adopted $P_{\rm min} = 1.2$~d and $P_{\rm max} = 10$~d following
\citet{Howard2012} and \citet{Fressin2013}.  We determined that among
1728 stars with no detected signals within the range of $1.2 < P < 10$
and $3 < R_p < 8$, assuming $\alpha = 1.9$, then $\displaystyle
\sum_i^{ND} \langle d_i \rangle = 16.8$.  This is the expected number
of detections around these stars if each had one such planet.  It is
not sensitive to the precise value of $\alpha$. 

\begin{figure}
\includegraphics[width=84mm]{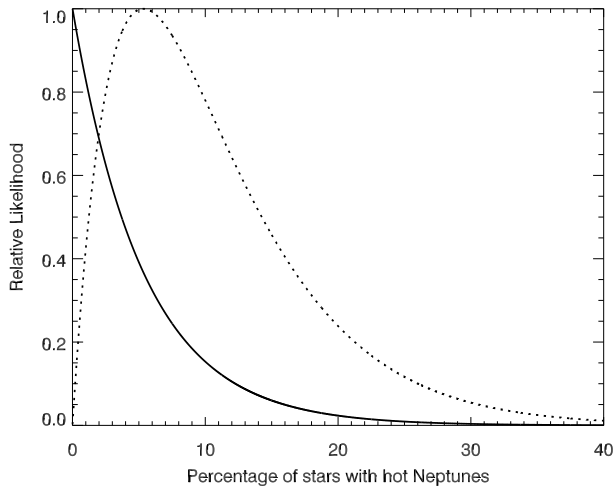}
\caption{Likelihood vs. occurrence of planets with 1.2~d $< P < 10$~d
  and 3\rearth{}$ < R_p < $8\rearth{} around SEAWOLF stars.  The
  dashed line is for the case of one confirmed detection and the solid
  line is for the case of no confirmed detections.}
\label{fig.occur}
\end{figure}

We calculated the likelihood vs. occurrence rate using Eqns,
\ref{eqn.occurrence}, \ref{eqn.geofactor}, and 10b.  Excluding
GJ~436b, and given that we have as yet no confirmed detections of new
planets in our sample, we can only place an upper limit on the
occurrence of hot Neptunes.  In this case, we place a 95\% confidence
upper limit of 10.2\% on $f$ based on a log likelihood within 1.92 of
the maximum value (Fig. \ref{fig.occur}).  We also estimate the most
likely occurrence in the case of a single confirmed detection: $f
=5.3\pm4.4$\% (Fig. \ref{fig.occur}).  The error is based on the
assumption of asymptopic normality; a parabola was iteratively fitted
to the log-likelihood curve and $\sigma_f = 1/\sqrt{2c}$, where $c$ is
the curvature of the parabola.

If there is more than one confirmed planet in our sample, the maximum
likelihood estimate of $f$ will be likewise higher.  If we relaxed the
assumption that all uncomfirmed detections are ruled out, then $f$
could be significantly higher, and close to unity, because the number
of WASP candidates that we have yet to screen is comparable to the
expected number ($\sim 17$) if every star had a hot Neptune.  However,
if our follow-up results are representative of the results as a whole,
then it is more likely that all or nearly all of these unscreened
systems will be ruled out as well.

\subsection{Comparison with \emph{Kepler}} \label{sec.kepler}

We estimated the occurrence of 3--8\rearth{} and $P < 10$~d planets
around \emph{Kepler} target stars using the January 2013 release of
confirmed and candidate transiting planets (KOIs) from analysis of
observation quarters Q1--Q8.  The methods are described in
\citet{Gaidos2013c} and here we recapitulate only the most important
details.  To emulate the range of spectral types of the SEAWOLF
survey, stars with $2 < V-J < 4.7$, with $V$ magnitudes based on the
relation $V = r + 0.44(g-r) - 0.02$ \citep{Fukugita1996}, were
selected from the complete \emph{Kepler} target catalog.  We also
required $K_p < 16$ and that each star was observed for at least seven
of the first eight observing quarters.  We estimated parameters for
these 14,578 stars and 190 (candidate) planets by fitting Dartmouth
stellar evolution models \citep{Dotter2008} using the Bayesian
procedure described in \citet{Gaidos2013b}.  We then limited the
analysis to 6422 stars with estimated $\log g > 4$ and $g$-D51 $<
0.23$.  D51 is an AB magnitude based on a passband centered on 510~nm
and the $g$-D51 color is an indicator of gravity among K dwarfs; the
color-cut eliminates K giants \citep{Brown2011AJ142}.  The median estimated
\teff{} of these stars is 4330~K.  These stars host 136 candidate
planets.  Two of these have 3\rearth{}$ < R_p < 8$\rearth{} and $P <
10$~d: KOIs 875.01 and 956.01 with $R_p$ of 3.7 and 3.2\rearth{} on
4.22 and 8.36~d orbits, respectively.

We calculated the binomial log likelihood as a function of
planet-hosting fraction $f$ assuming a log distribution with orbital
period and a power-law radius distribution in the limit that the
transit probability is low \citep{Mann2012,Gaidos2013c}:
\begin{equation}
\label{eqn.f}
f = \frac{N_p\left(R_1^{-\alpha}-R_2^{-\alpha}\right)\ln \left(P_2/P_1\right)}{\alpha \displaystyle \sum_i^{ND}\langle f_i \rangle}
\end{equation}
where $N_p$ is the number of detected planets with $R_1 < R_p < R_2$ and $P_1 < P < P_2$, the summation in the denominator is over non-detections, 
\begin{equation}
\label{eqn.f2}
\langle f_i  \rangle = \int_{R_1}^{R_2}\int_{P_1}^{P_2} R_p^{-\alpha}d_i(R_p,P) \, d\ln P d\ln R_p,
\end{equation}
and $d_i(R_p,P)$ is the probability of detecting a planet around the
$i$th star \citep{Mann2012}.  For consistency with SEAWOLF we use $P_1
= 1.2$~d, $P_2 = 10$~d, and $\alpha = 1.9$.

The transit of a late K or early M dwarf by a Neptune-size planet
produces a signal of magnitude $4 \times 10^{-3}$, far larger than the
noise: The median 3~hr Combined Differential Photometry Precision
(CDPP3) for the stars in our sample is $1.8 \times 10^{-4}$ and the 99
percentile value is $6.6\times 10^{-4}$.  We estimated the cumulative
SNR from a 3\rearth{} planet on a 10~d orbit monitored for 2~yr (8
quarters): this is the least detectable case.  The stellar noise over
the transit interval was taken to be the CDPP3 scaled by
$\sqrt{3/\tau}$ where $\tau$ is the transit duration in hours.
\citet{Fressin2013} found that the recovery rate of the \emph{Kepler}
detection pipeline is nearly 100\% for SNR$>$16.  Of the 9741 stars
with CDPP3 values, for only 33 (0.3\%) would the estimated SNR be
$<16$.

Thus the detection probability is essentially the geometric factor
$R_*/a$, where $a$ is the orbital semimajor axis, and independent of
$R_p$.  The $R_p$ terms in Eqn. \ref{eqn.f} cancel and
\begin{equation}
f = N_p \ln (P_2/P_1)/\displaystyle \sum_i^{ND}F_i.
\end{equation}
The detection probability becomes:
\begin{equation}
d_i(P) = \left(\frac{4\pi^2 R_*^3}{GM_*}\right)^{1/3}\frac{1 + e \cos \omega}{1-e^2} P^{-2/3},
\end{equation}
where $e$ is the orbital eccentricity and $\omega$ the longitude of
periastron.  Marginalizing over $e$ and $\omega$ with an eccentricity
distribution $n(e)$, and ignoring terms that do not depend on $f$,
\begin{equation}
\label{eqn.likelihood2}
\begin{split}
\ln \mathcal{L} \approx N_D \ln f - 0.356 f \left[\int_0^1\frac{n(e)de}{1-e^2}\right]\left(\frac{P_2}{\rm 1~d}\right)^{-2/3} \\ \times \frac{\left(P_2/P_1\right)^{2/3}-1}{\ln (P_2/P_1)}\displaystyle \sum_j^{ND}\left(\frac{\rho_j}{\rho_{\odot}}\right)^{-1/3}+ \cdots, 
\end{split}
\end{equation}
where $N_D$ is the number of detected planets and $\rho$ is the mean
stellar density.  Adopting the function for $n(e)$ in
\citet{Shen2008}, we found that the integral is only weakly dependent
on the parameter $a$ in their distribution, and is $\approx 1.20$ for
$a=4$.  Using a Rayleigh distribution like that in \citet{Gaidos2013c}
gives a similar value of 1.08 for the integral.  Because each star can
be explained by more than one stellar model with probability $p$, we
used a weighted mean of $\rho^{-1/3}$ to calculate the likelihood:
\begin{equation}
\langle \rho^{-1/3} \rangle = \displaystyle \sum_i p_i \rho_i^{-1/3} /\displaystyle \sum_i p_i,
\end{equation}
where the summation is restricted to main sequence models, i.e. $\log
g > 4$.  

\begin{figure}
\includegraphics[width=84mm]{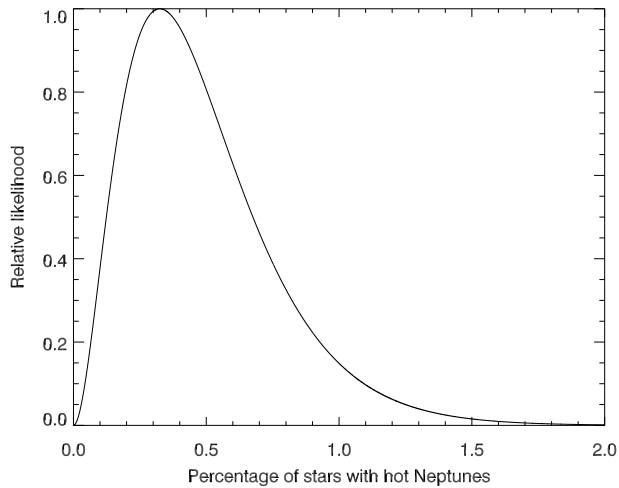}
\caption{Likelihood vs. occurrence of planets with $P < 10$~d and
  3\rearth{}$ < R_p < $8\rearth{} around 6422 dwarf stars wtih $2 <
  V-J < 4.7$ observed by \emph{Kepler} during at least 7 quarters of
  Q1-8.} \label{fig.kepler}
\end{figure}

Under these assumptions, we found that the occurrence of hot ($P <
10$~d) Neptunes is $0.33 \pm 0.21$\% (Fig. \ref{fig.kepler}).

\section[]{Discussion and Conclusions} \label{sec.discussion}

We place a limit of 10\% on the occurence of hot Neptunes ($P < 10$~d)
around the late K and early M dwarfs in our SEAWOLF sample (95\%
confidence).  In the event that a single planet candidate is
confirmed, our maximum likelihood estimate of occurrence is $5.3 \pm
4.4$\%.  From a Doppler survey of late F to early K stars,
\citet{Howard2010} estimated an occurrence rate of about $8.1 \pm 4$\%
for planets with projected masses $M \sin i$ of 10-100\mearth{}, a
mass range correspondingly approximately to our radius range, and $P <
50$~d.  Assuming a logarithmic distribution with orbital period, and
correcting by the factor $\ln (10/1.2)/\ln (50/1.2)$, the equivalent
occurrence within 10~d is 4.6\%, a value similar to our estimate for
the case of a single detection.  Based on the HARPS Doppler survey,
\citet{Mayor2011} estimated $11.1\pm2.4$\% of solar-type stars have
planets with 10-30\mearth{} within $P < 50$~d, but only
$1.17\pm0.52$\% with masses of 30--100\mearth{}.  Likewise,
\citet{Bonfils2013} estimate that $3^{+4}_{-1}$\% of M dwarfs have
10-100\mearth planets.  That these Doppler-based values are consistent
with the 5.3\% occurrence we derive assuming a single SEAWOLF
detection suggests that GJ~436b may be the only transiting hot Neptune
in our sample.

We estimated the occurrence of hot Neptunes around the late K and
early M dwarf stars observed by \emph{Kepler} to be $0.32 \pm 0.21$\%,
more than an order of magnitude lower than in the SEAWOLF
catalog. \citet{Howard2012} report that the occurrence of
4--8\rearth{} planets with $P < 10$~d around \emph{Kepler} GK dwarfs
is $0.23 \pm 0.03$\%, consistent with our estimate to within the
errors (but for a different range of spectral types).  One major
caveat with interpreting the \emph{Kepler} statistics is that late K
spectral types also include red giant branch as well as dwarf stars;
these luminosity classes can be difficult to distinguish by
photometric colors alone and in the absence of spectroscopic
screening, Malmquist bias will favor the inclusion of the large, more
luminous stars \citep{Gaidos2013}.  Planets will be more difficult to
detect and will appear smaller around RGB stars, e.g. some ``Earths''
may actually be Neptunes.  For this reason, it is possible that the
statistical analysis of the \emph{Kepler} results grossly
underestimates the occurrence of hot Neptunes.  However, our use of
the $g$-D51 gravity-sensitive color in constructing our \emph{Kepler}
sample should limit this effect.  More spectroscopy of \emph{Kepler}
targets in this range of $V-J$ colors is needed to quantify
contamination by RGB stars.

Our determination of an order-of-magnitude lower relative occurrence
of hot Neptunes around \emph{Kepler} stars echoes the findings of
\citet{Wright2012}, who found a deficit of hot Jupiters around these
stars.  One intriguing possibility is that \emph{Kepler} stars are
older, more evolved, and have more massive convective envelopes than
those in the Solar neighborhood, and that close-in giant planets have
suffered tidally-driven decay of their orbits and been destroyed
\citep{Gaidos2013}.  However, in the regime where the orbital period
$P$ is much shorter than the eddy turnover timescale $T$, the rate of
orbital decay is \citep{Kunitomo2011}:
\begin{equation}
\label{eqn.decaytime}
\frac{\dot{a}}{a} = \frac{3}{4}\frac{M_p}{M_*}\frac{L_*P^2}{M_*\left(R_*-R_{\rm env}\right)^2}\left(\frac{R_*}{a}\right)^8,
\end{equation}
where $M_p$ is the mass of the planet, $L_*$ is the stellar
luminosity, $R_{\rm env}$ the inner radius of the convective envelope,
and $a$ the orbital semimajor axis of the planet.  For a Neptune-mass
planet on a 10~d orbit around an M0 dwarf star the theoretical orbital
decay time is $>10^{18}$~yr.  The lower luminosity and smaller radius
of K/M dwarfs relative to solar-type stars, and the lower mass of
Neptunes relative to Jupiters means that this process is too slow to
explain the deficit of Neptunes close to \emph{Kepler} stars relative
to the solar neighborhood.  \emph{Kepler} stars may also be more
metal-poor than the solar enighborhood, but there is, as yet, no
evidence that the occurrence of Neptunes depnds on the metallicity of
the host star \citep{Mann2013b}.  Instead, the discrepancy must be
explained by observational selection or differences in the efficiency
in formation on or migration to close-in orbits.

The Next Generation Transit Survey \citep{Wheatley2013} will monitor
$\sim 1.6 \times 10^6$ K4-M4 dwarfs with $I< 17$ to search for hot
Neptunes; about $2 \times 10^{5}$ of these stars will have $I < 15$
and are suitable for Doppler follow-up (P. Wheatley, personal
communication).  If our 0.32\% occurrence rate from \emph{Kepler} is
correct the number of transiting hot (super)Neptunes in this survey
will be $\le 60$, depending on actual detection efficiency.  If the
the occurrence rate is close to 5\%, as suggested by Doppler surveys
of nearby stars, the survey could find up to $\sim 1000$ such planets.
These two values bracket an estimate by the NGTS team
\citep{Wheatley2013}.

\section*{Acknowledgments}

EG and KC acknowledge support from NASA grants NNX10AI90G
(Astrobiology: Exobiology \& Evolutionary Biology) and NNX11AC33G
(Origins of Solar Sytems).  GM acknowledges support from the State
Scholarships Foundation of Greece in the form of a scholarship and
fruitful discussions with E. Paleologou.  AZ acknowledges the
Foundation for Research and Technology-Hellas.  Space astrophysics in
Crete is supported in part by EU REGPOT project number 206469.  NN
acknowledges support by an NAOJ Fellowship, NINS Program for
Cross-Disciplinary Study, and Grant-in-Aid for Scientific Research (A)
(No.~25247026) from the Ministry of Education, Culture, Sports,
Science and Technology of Japan.  EC and JB were supported by
NASA/University of Hawaii Space Grant fellowships.  The SuperWASP-N
Camera was constructed and operated with funds made available from
WASP Consortium universities and the UK Science and Technology
Facilities Council (STFC). We extend our thanks to the Director and
staff of the Isaac Newton Group of Telescopes for their support of
SuperWASP-N operations.  We thank the staff of the MDM and Skinakas
observatories for their invaluable and courteous assistance during
many observing runs.  The Las Cumbres Observatory Network operates
Faulkes Telescope North, ELP at McDonald Observatory, and the Byrne
Observatory; the last is located on the Sedgwick Reserve, a part of
the University of California Natural Reserve System.

\include{table1}

\include{table2_short}

\include{table3_short}

\label{lastpage}

\end{document}

%% file: table1.tex
\begin{table*}
 \centering
 \begin{minipage}{140mm}
  \caption{Telescopes used to obtain follow-up observations \label{tab.telescopes}}
  \begin{tabular}{@{}llrlllr@{}}
    \hline
    \multicolumn{1}{c}{Telescope/Observatory} & \multicolumn{1}{c}{D (m)} & \multicolumn{1}{c}{Latitude} & \multicolumn{1}{c}{Longitude} & \multicolumn{1}{c}{Instrument(s)} & \multicolumn{1}{c}{Passband(s)} & \multicolumn{1}{c}{Observations\footnote{Usable data of a candidate transit event ($C > 0$)}}\\
    \hline
    McGraw-Hill/MDM & 1.3 & 31.95173 N & 111.61664 W & B4K/R4K/Nellie & Sloan $r$, DES-$Z$ & 66\\
    Faulkes North/Haleakala & 2.0 & 20.70701 N & 156.25748 W & SpectraCam 4K & Pan-STARRS $Z$ & 29\\
    Skinakas & 1.3 & 35.21173 N & 024.89893 E & Andor DZ436 & Bessel $R$ & 24\\
    OAO 188~cm & 1.88 & 34.57716 N & 133.59387 E & ISLE\footnote{\citet{Yanagisawa2006}} & $J$ & 19 \\
    LCOGT/BOS & 0.8 & 34.68750 N & 120.03889 W & SBIG & Sloan $i$' & 5\\
    LCOGT/ELP & 1.0 & 30.67143 N & 104.02195 W & kb73 & Sloan $i$ & 4\\
    UH88/Mauna Kea & 2.2 & 19.82303 N & 155.46937 W & OPTIC & Sloan $z$ & 1\\
    \hline
  \end{tabular}
\end{minipage}
\end{table*}

%% file: table2_short.tex
\begin{table*}
\centering
\begin{minipage}{140mm}
\caption{Candidate Transit Systems Identified in WASP Data \label{tab.candidates}}
\begin{tabular}{@{}lrrrrrrrrrrrl@{}}
\hline
\multicolumn{1}{c}{Name\footnote{ABC refer to multiple signals for the same star}} & \multicolumn{3}{c}{RA} & \multicolumn{3}{c}{Dec} & \multicolumn{1}{c}{V} & \multicolumn{1}{c}{V-J} & \multicolumn{1}{c}{period} & \multicolumn{1}{c}{ephemeris} & \multicolumn{1}{c}{$\delta$} & \multicolumn{1}{c}{status\footnote{X = ruled out, ? = ambiguous or insufficient data, A = candidate, N = not observed}} \\
 & hh & mm & ss.s & dd & mm & ss &  &  & \multicolumn{1}{c}{(d)} & \multicolumn{1}{c}{(BJD)} & ($10^{-3}$) &  \\
\hline
00177+2100  &  0 & 17 & 43.2 &  21 & 00 & 05 & 12.4 & 2.45 &  5.319 &  3878.1780 &     4.6 & ? \\
00492+2003  &  0 & 49 & 17.2 &  20 & 03 & 45 & 10.8 & 2.28 &  8.063 &  3502.8216 &     3.3 & ? \\
01086+1714A &  1 & 08 & 40.4 &  17 & 14 & 33 & 10.7 & 2.66 &  5.530 &  4177.7331 &     4.3 & X \\
01086+1714B &  1 & 08 & 40.4 &  17 & 14 & 33 & 10.7 & 2.66 &  3.743 &  4169.6813 &     3.2 & ? \\
01550+4035  &  1 & 55 & 01.0 &  40 & 35 & 06 & 13.7 & 3.91 &  8.343 &  3387.8129 &    20.3 & ? \\
01578+3130  &  1 & 57 & 50.0 &  31 & 30 & 41 & 12.2 & 2.25 &  4.177 &  4225.8173 &     4.1 & X \\
01587+3515  &  1 & 58 & 43.6 &  35 & 15 & 28 & 13.7 & 4.01 &  7.819 &  4191.1928 &     9.7 & X \\
02083+2919  &  2 & 08 & 18.3 &  29 & 19 & 59 & 12.5 & 2.93 &  7.214 &  3786.2246 &     8.0 & X \\
02111+2707  &  2 & 11 & 11.2 &  27 & 07 & 34 & 12.7 & 3.36 & 10.560 &  3894.2242 &     8.8 & N \\
02192+2456A &  2 & 19 & 17.5 &  24 & 56 & 38 & 13.8 & 4.09 &  6.985 &  4022.5748 &    11.9 & N \\
\hline
\end{tabular}
\medskip
Table \ref{tab.candidates} is published in its entirety as a machine-readable table in the CDS.  A portion is shown here for guidance regarding its form and content.
\end{minipage}
\end{table*}

%% file: table3_short.tex
\begin{table*}
\centering
\begin{minipage}{95mm}
\caption{Observations of Candidate Transits \label{tab.observations}}
\begin{tabular}{@{}lllrr@{}}
\hline
\multicolumn{1}{c}{Star\footnote{ABC refer to multiple signals for the same star}} & \multicolumn{1}{c}{Observatory} & \multicolumn{1}{c}{$t_c$ (MJD)} & \multicolumn{1}{c}{$\delta_d$} & \multicolumn{1}{c}{$C$} \\
 &  & \multicolumn{1}{c}{($-2.45 \times 10^6$)} & \multicolumn{1}{c}{($10^{-3}$)} &  \\
\hline
00177+2100  &           MDM & 6191.85 &  16.6 & 0.032 \\
00177+2100  &           MDM & 6207.80 &   8.7 & 0.090 \\
00492+2003  &           MDM & 6195.75 &   2.3 & 0.644 \\
01086+1714A & LCOGT/Faulkes & 6212.81 &   1.4 & 0.940 \\
01086+1714A &           MDM & 6284.71 &   6.4 & 0.206 \\
01086+1714B & LCOGT/Faulkes & 5898.79 &   7.5 & 0.047 \\
01086+1714B & LCOGT/Faulkes & 6157.03 &   3.9 & 0.360 \\
01086+1714B & LCOGT/Faulkes & 6564.98 &   7.2 & 0.061 \\
01086+1714B &           MDM & 6583.69 &   5.1 & 0.165 \\
01550+4035  &     LCOGT/BOS & 6524.85 &   1.6 & 0.199 \\
\hline
\end{tabular}
\medskip
Table \ref{tab.observations} is published in its entirety as a machine-readable table in the CDS. A portion is shown here for guidance regarding its form and content.
\end{minipage}
\end{table*}

%% file: manuscript_main.bbl
\begin{thebibliography}{61}
\expandafter\ifx\csname natexlab\endcsname\relax\def\natexlab#1{#1}\fi

\bibitem[{{Bakos} {et~al.}(2011){Bakos}, {Hartman}, {Torres}, {Kov{\'a}cs},
  {Noyes}, {Latham}, {Sasselov}, \& {B{\'e}ky}}]{Bakos2011}
{Bakos}, G.~{\'A}., {Hartman}, J.~D., {Torres}, G., {Kov{\'a}cs}, G., {Noyes},
  R.~W., {Latham}, D.~W., {Sasselov}, D.~D., \& {B{\'e}ky}, B. 2011, in
  European Physical Journal Web of Conferences, Vol.~11, Detection and Dynamics
  of Transiting Exoplanets, 1002

\bibitem[{{Baraffe} {et~al.}(2005){Baraffe}, {Chabrier}, {Barman}, {Selsis},
  {Allard}, \& {Hauschildt}}]{Baraffe2005}
{Baraffe}, I., {Chabrier}, G., {Barman}, T.~S., {Selsis}, F., {Allard}, F., \&
  {Hauschildt}, P.~H. 2005, \aap, 436, L47

\bibitem[{{Berta} {et~al.}(2012){Berta}, {Irwin}, {Charbonneau}, {Burke}, \&
  {Falco}}]{Berta2012}
{Berta}, Z.~K., {Irwin}, J., {Charbonneau}, D., {Burke}, C.~J., \& {Falco},
  E.~E. 2012, \aj, 144, 145

\bibitem[{{Bonfils} {et~al.}(2012){Bonfils}, {Gillon}, {Udry}, {Armstrong},
  {Bouchy}, {Delfosse}, {Forveille}, {Fumel}, {Jehin}, {Lendl}, {Lovis},
  {Mayor}, {McCormac}, {Neves}, {Pepe}, {Perrier}, {Pollaco}, {Queloz}, \&
  {Santos}}]{Bonfils2012}
{Bonfils}, X., {et~al.} 2012, \aap, 546, A27

\bibitem[{{Bonfils} {et~al.}(2013){Bonfils}, {Delfosse}, {Udry}, {Forveille},
  {Mayor}, {Perrier}, {Bouchy}, {Gillon}, {Lovis}, {Pepe}, {Queloz}, {Santos},
  {S{\'e}gransan}, \& {Bertaux}}]{Bonfils2013}
---. 2013, \aap, 549, A109

\bibitem[{{Borucki} {et~al.}(2010){Borucki}, {Koch}, {Basri}, {Batalha},
  {Brown}, {Caldwell}, {Caldwell}, {Christensen-Dalsgaard}, {Cochran},
  {DeVore}, {Dunham}, {Dupree}, {Gautier}, {Geary}, {Gilliland}, {Gould},
  {Howell}, {Jenkins}, {Kondo}, {Latham}, {Marcy}, {Meibom}, {Kjeldsen},
  {Lissauer}, {Monet}, {Morrison}, {Sasselov}, {Tarter}, {Boss}, {Brownlee},
  {Owen}, {Buzasi}, {Charbonneau}, {Doyle}, {Fortney}, {Ford}, {Holman},
  {Seager}, {Steffen}, {Welsh}, {Rowe}, {Anderson}, {Buchhave}, {Ciardi},
  {Walkowicz}, {Sherry}, {Horch}, {Isaacson}, {Everett}, {Fischer}, {Torres},
  {Johnson}, {Endl}, {MacQueen}, {Bryson}, {Dotson}, {Haas}, {Kolodziejczak},
  {Van Cleve}, {Chandrasekaran}, {Twicken}, {Quintana}, {Clarke}, {Allen},
  {Li}, {Wu}, {Tenenbaum}, {Verner}, {Bruhweiler}, {Barnes}, \&
  {Prsa}}]{Borucki2010Sci}
{Borucki}, W.~J., {et~al.} 2010, Science, 327, 977

\bibitem[{{Bou{\'e}} {et~al.}(2012){Bou{\'e}}, {Figueira}, {Correia}, \&
  {Santos}}]{Boue2012}
{Bou{\'e}}, G., {Figueira}, P., {Correia}, A.~C.~M., \& {Santos}, N.~C. 2012,
  \aap, 537, L3

\bibitem[{{Boyajian} {et~al.}(2012){Boyajian}, {von Braun}, {van Belle},
  {McAlister}, {ten Brummelaar}, {Kane}, {Muirhead}, {Jones}, {White},
  {Schaefer}, {Ciardi}, {Henry}, {L{\'o}pez-Morales}, {Ridgway}, {Gies}, {Jao},
  {Rojas-Ayala}, {Parks}, {Sturmann}, {Sturmann}, {Turner}, {Farrington},
  {Goldfinger}, \& {Berger}}]{Boyajian2012ApJ757}
{Boyajian}, T.~S., {et~al.} 2012, \apj, 757, 112

\bibitem[{{Brown} {et~al.}(2011){Brown}, {Latham}, {Everett}, \&
  {Esquerdo}}]{Brown2011AJ142}
{Brown}, T.~M., {Latham}, D.~W., {Everett}, M.~E., \& {Esquerdo}, G.~A. 2011,
  \aj, 142, 112

\bibitem[{{Brunini} \& {Cionco}(2005)}]{Brunini2005}
{Brunini}, A., \& {Cionco}, R.~G. 2005, \icarus, 177, 264

\bibitem[{{Carone} {et~al.}(2012){Carone}, {Gandolfi}, {Cabrera}, {Hatzes},
  {Deeg}, {Csizmadia}, {P{\"a}tzold}, {Weingrill}, {Aigrain}, {Alonso},
  {Alapini}, {Almenara}, {Auvergne}, {Baglin}, {Barge}, {Bonomo}, {Bord{\'e}},
  {Bouchy}, {Bruntt}, {Carpano}, {Cochran}, {Deleuil}, {D{\'{\i}}az},
  {Dreizler}, {Dvorak}, {Eisl{\"o}ffel}, {Eigm{\"u}ller}, {Endl}, {Erikson},
  {Ferraz-Mello}, {Fridlund}, {Gazzano}, {Gibson}, {Gillon}, {Gondoin},
  {Grziwa}, {G{\"u}nther}, {Guillot}, {Hartmann}, {Havel}, {H{\'e}brard},
  {Jorda}, {Kabath}, {L{\'e}ger}, {Llebaria}, {Lammer}, {Lovis}, {MacQueen},
  {Mayor}, {Mazeh}, {Moutou}, {Nortmann}, {Ofir}, {Ollivier}, {Parviainen},
  {Pepe}, {Pont}, {Queloz}, {Rabus}, {Rauer}, {R{\'e}gulo}, {Renner}, {de La
  Reza}, {Rouan}, {Santerne}, {Samuel}, {Schneider}, {Shporer}, {Stecklum},
  {Tal-Or}, {Tingley}, {Udry}, \& {Wuchterl}}]{Carone2012}
{Carone}, L., {et~al.} 2012, \aap, 538, A112

\bibitem[{Charbonneau \& Brown(2000)}]{Charbonneau2000}
Charbonneau, D., \& Brown, T. 2000, \apj, 529, L45

\bibitem[{Charbonneau {et~al.}(2009)Charbonneau, Berta, Irwin, Burke, Nutzman,
  Buchhave, Lovis, Bonfils, Latham, Udry, Murray-Clay, Holman, Falco, Winn,
  Queloz, Pepe, Mayor, Delfosse, \& Forveille}]{Charbonneau2009a}
Charbonneau, D., {et~al.} 2009, Nature, 462, 891

\bibitem[{{Christian} {et~al.}(2006){Christian}, {Pollacco}, {Skillen},
  {Street}, {Keenan}, {Clarkson}, {Collier Cameron}, {Kane}, {Lister}, {West},
  {Enoch}, {Evans}, {Fitzsimmons}, {Haswell}, {Hellier}, {Hodgkin}, {Horne},
  {Irwin}, {Norton}, {Osborne}, {Ryans}, {Wheatley}, \&
  {Wilson}}]{Christian2006}
{Christian}, D.~J., {et~al.} 2006, \mnras, 372, 1117

\bibitem[{{Claret}(2000)}]{Claret2000}
{Claret}, A. 2000, \aap, 363, 1081

\bibitem[{{Claret}(2004)}]{Claret2004}
---. 2004, \aap, 428, 1001

\bibitem[{{Collier Cameron} {et~al.}(2006){Collier Cameron}, {Pollacco},
  {Street}, {Lister}, {West}, {Wilson}, {Pont}, {Christian}, {Clarkson},
  {Enoch}, {Evans}, {Fitzsimmons}, {Haswell}, {Hellier}, {Hodgkin}, {Horne},
  {Irwin}, {Kane}, {Keenan}, {Norton}, {Parley}, {Osborne}, {Ryans}, {Skillen},
  \& {Wheatley}}]{CollierCameron2006}
{Collier Cameron}, A., {et~al.} 2006, \mnras, 373, 799

\bibitem[{Cox(2000)}]{Cox2000}
Cox, A.~N. 2000, {Allen's Astrophysical Quantities} (New York: AIP Press,
  Springer)

\bibitem[{{Cumming} {et~al.}(2008){Cumming}, {Butler}, {Marcy}, {Vogt},
  {Wright}, \& {Fischer}}]{Cumming2008}
{Cumming}, A., {Butler}, R.~P., {Marcy}, G.~W., {Vogt}, S.~S., {Wright}, J.~T.,
  \& {Fischer}, D.~A. 2008, \pasp, 120, 531

\bibitem[{{Dotter} {et~al.}(2008){Dotter}, {Chaboyer}, {Jevremovi{\'c}},
  {Kostov}, {Baron}, \& {Ferguson}}]{Dotter2008}
{Dotter}, A., {Chaboyer}, B., {Jevremovi{\'c}}, D., {Kostov}, V., {Baron}, E.,
  \& {Ferguson}, J.~W. 2008, \apjs, 178, 89

\bibitem[{{Duncan} {et~al.}(1991){Duncan}, {Vaughan}, {Wilson}, {Preston},
  {Frazer}, {Lanning}, {Misch}, {Mueller}, {Soyumer}, {Woodard}, {Baliunas},
  {Noyes}, {Hartmann}, {Porter}, {Zwaan}, {Middelkoop}, {Rutten}, \&
  {Mihalas}}]{Duncan1991}
{Duncan}, D.~K., {et~al.} 1991, \apjs, 76, 383

\bibitem[{{Fressin} {et~al.}(2013){Fressin}, {Torres}, {Charbonneau}, {Bryson},
  {Christiansen}, {Dressing}, {Jenkins}, {Walkowicz}, \&
  {Batalha}}]{Fressin2013}
{Fressin}, F., {et~al.} 2013, \apj, 766, 81

\bibitem[{{Fukugita} {et~al.}(1996){Fukugita}, {Ichikawa}, {Gunn}, {Doi},
  {Shimasaku}, \& {Schneider}}]{Fukugita1996}
{Fukugita}, M., {Ichikawa}, T., {Gunn}, J.~E., {Doi}, M., {Shimasaku}, K., \&
  {Schneider}, D.~P. 1996, \aj, 111, 1748

\bibitem[{{Gaidos}(2013)}]{Gaidos2013b}
{Gaidos}, E. 2013, \apj, 770, 90

\bibitem[{{Gaidos} {et~al.}(2013){Gaidos}, {Fischer}, {Mann}, \&
  {Howard}}]{Gaidos2013c}
{Gaidos}, E., {Fischer}, D.~A., {Mann}, A.~W., \& {Howard}, A.~W. 2013, \apj,
  771, 18

\bibitem[{{Gaidos} \& {Mann}(2013)}]{Gaidos2013}
{Gaidos}, E., \& {Mann}, A.~W. 2013, \apj, 762, 41

\bibitem[{Gillon {et~al.}(2007)Gillon, Pont, Demory, Mallmann, Mayor, Mazeh,
  Queloz, Shporer, Udry, \& Vuissoz}]{Gillon2007}
Gillon, M., {et~al.} 2007, \aap, 472, L13

\bibitem[{{Hansen} \& {Murray}(2012)}]{Hansen2012}
{Hansen}, B.~M.~S., \& {Murray}, N. 2012, \apj, 751, 158

\bibitem[{{Hartman} {et~al.}(2011){Hartman}, {Bakos}, {Kipping}, {Torres},
  {Kov{\'a}cs}, {Noyes}, {Latham}, {Howard}, {Fischer}, {Johnson}, {Marcy},
  {Isaacson}, {Quinn}, {Buchhave}, {B{\'e}ky}, {Sasselov}, {Stefanik},
  {Esquerdo}, {Everett}, {Perumpilly}, {L{\'a}z{\'a}r}, {Papp}, \&
  {S{\'a}ri}}]{Hartman2011}
{Hartman}, J.~D., {et~al.} 2011, \apj, 728, 138

\bibitem[{Henry {et~al.}(2000)Henry, Marcy, Butler, \& Vogt}]{Henry2000}
Henry, G.~W., Marcy, G.~W., Butler, R.~P., \& Vogt, S.~S. 2000, \apj, 529, L41

\bibitem[{{Howard} {et~al.}(2010){Howard}, {Marcy}, {Johnson}, {Fischer},
  {Wright}, {Isaacson}, {Valenti}, {Anderson}, {Lin}, \& {Ida}}]{Howard2010}
{Howard}, A.~W., {et~al.} 2010, Science, 330, 653

\bibitem[{{Howard} {et~al.}(2012){Howard}, {Marcy}, {Bryson}, {Jenkins},
  {Rowe}, {Batalha}, {Borucki}, {Koch}, {Dunham}, {Gautier}, {Van Cleve},
  {Cochran}, {Latham}, {Lissauer}, {Torres}, {Brown}, {Gilliland}, {Buchhave},
  {Caldwell}, {Christensen-Dalsgaard}, {Ciardi}, {Fressin}, {Haas}, {Howell},
  {Kjeldsen}, {Seager}, {Rogers}, {Sasselov}, {Steffen}, {Basri},
  {Charbonneau}, {Christiansen}, {Clarke}, {Dupree}, {Fabrycky}, {Fischer},
  {Ford}, {Fortney}, {Tarter}, {Girouard}, {Holman}, {Johnson}, {Klaus},
  {Machalek}, {Moorhead}, {Morehead}, {Ragozzine}, {Tenenbaum}, {Twicken},
  {Quinn}, {Isaacson}, {Shporer}, {Lucas}, {Walkowicz}, {Welsh}, {Boss},
  {Devore}, {Gould}, {Smith}, {Morris}, {Prsa}, {Morton}, {Still}, {Thompson},
  {Mullally}, {Endl}, \& {MacQueen}}]{Howard2012}
---. 2012, \apjs, 201, 15

\bibitem[{{H{\"u}nsch} {et~al.}(1999){H{\"u}nsch}, {Schmitt}, {Sterzik}, \&
  {Voges}}]{Hunsch1999}
{H{\"u}nsch}, M., {Schmitt}, J.~H.~M.~M., {Sterzik}, M.~F., \& {Voges}, W.
  1999, \aaps, 135, 319

\bibitem[{{Isaacson} \& {Fischer}(2010)}]{Isaacson2010}
{Isaacson}, H., \& {Fischer}, D. 2010, \apj, 725, 875

\bibitem[{{Kov{\'a}cs} {et~al.}(2005){Kov{\'a}cs}, {Bakos}, \&
  {Noyes}}]{Kovacs2005}
{Kov{\'a}cs}, G., {Bakos}, G., \& {Noyes}, R.~W. 2005, \mnras, 356, 557

\bibitem[{{Kov{\'a}cs} {et~al.}(2002){Kov{\'a}cs}, {Zucker}, \&
  {Mazeh}}]{Kovacs2002}
{Kov{\'a}cs}, G., {Zucker}, S., \& {Mazeh}, T. 2002, \aap, 391, 369

\bibitem[{Kunitomo {et~al.}(2011)Kunitomo, Ikoma, Sato, Katsuta, \&
  Ida}]{Kunitomo2011}
Kunitomo, M., Ikoma, M., Sato, B., Katsuta, Y., \& Ida, S. 2011, Astrophys. J.,
  737, 66

\bibitem[{{Laughlin} {et~al.}(2004){Laughlin}, {Bodenheimer}, \&
  {Adams}}]{Laughlin2004}
{Laughlin}, G., {Bodenheimer}, P., \& {Adams}, F.~C. 2004, \apjl, 612, L73

\bibitem[{L\'{e}pine \& Gaidos(2011)}]{Lepine2011}
L\'{e}pine, S., \& Gaidos, E. 2011, \aj, 142, 138

\bibitem[{Lepine \& Shara(2005)}]{Lepine2005}
Lepine, S., \& Shara, M.~M. 2005, \aj, 129, 1483

\bibitem[{{Mann} {et~al.}(2011){Mann}, {Gaidos}, \& {Aldering}}]{Mann2011}
{Mann}, A.~W., {Gaidos}, E., \& {Aldering}, G. 2011, \pasp, 123, 1273

\bibitem[{{Mann} {et~al.}(2013){Mann}, {Gaidos}, {Kraus}, \&
  {Hilton}}]{Mann2013b}
{Mann}, A.~W., {Gaidos}, E., {Kraus}, A., \& {Hilton}, E.~J. 2013, \apj, 770,
  43

\bibitem[{{Mann} {et~al.}(2012){Mann}, {Gaidos}, {L{\'e}pine}, \&
  {Hilton}}]{Mann2012}
{Mann}, A.~W., {Gaidos}, E., {L{\'e}pine}, S., \& {Hilton}, E.~J. 2012, \apj,
  753, 90

\bibitem[{{Markwardt}(2009)}]{Markwardt2009}
{Markwardt}, C.~B. 2009, in ASP Conference Series, Vol. 411, Astronomical Data
  Analysis Software and Systems XVIII, ed. D.~A. {Bohlender}, D.~{Durand}, \&
  P.~{Dowler}, 251

\bibitem[{{Mayor} {et~al.}(2011){Mayor}, {Marmier}, {Lovis}, {Udry},
  {S{\'e}gransan}, {Pepe}, {Benz}, {Bertaux}, {Bouchy}, {Dumusque}, {Lo Curto},
  {Mordasini}, {Queloz}, \& {Santos}}]{Mayor2011}
{Mayor}, M., {et~al.} 2011, ArXiv e-prints

\bibitem[{{McCook} \& {Sion}(1987)}]{McCook1987}
{McCook}, G.~P., \& {Sion}, E.~M. 1987, \apjs, 65, 603

\bibitem[{{McNeil} \& {Nelson}(2010)}]{McNeil2010}
{McNeil}, D.~S., \& {Nelson}, R.~P. 2010, \mnras, 401, 1691

\bibitem[{{Mordasini} {et~al.}(2009){Mordasini}, {Alibert}, \&
  {Benz}}]{Mordasini2009}
{Mordasini}, C., {Alibert}, Y., \& {Benz}, W. 2009, \aap, 501, 1139

\bibitem[{{Pollacco} {et~al.}(2006){Pollacco}, {Skillen}, {Collier Cameron},
  {Christian}, {Hellier}, {Irwin}, {Lister}, {Street}, {West}, {Anderson},
  {Clarkson}, {Deeg}, {Enoch}, {Evans}, {Fitzsimmons}, {Haswell}, {Hodgkin},
  {Horne}, {Kane}, {Keenan}, {Maxted}, {Norton}, {Osborne}, {Parley}, {Ryans},
  {Smalley}, {Wheatley}, \& {Wilson}}]{Pollacco2006}
{Pollacco}, D.~L., {et~al.} 2006, \pasp, 118, 1407

\bibitem[{{Pont} {et~al.}(2006){Pont}, {Zucker}, \& {Queloz}}]{Pont2006}
{Pont}, F., {Zucker}, S., \& {Queloz}, D. 2006, \mnras, 373, 231

\bibitem[{{Reid} {et~al.}(1995){Reid}, {Hawley}, \& {Gizis}}]{Reid1995}
{Reid}, I.~N., {Hawley}, S.~L., \& {Gizis}, J.~E. 1995, \aj, 110, 1838

\bibitem[{{Rogers} {et~al.}(2011){Rogers}, {Bodenheimer}, {Lissauer}, \&
  {Seager}}]{Rogers2011}
{Rogers}, L.~A., {Bodenheimer}, P., {Lissauer}, J.~J., \& {Seager}, S. 2011,
  \apj, 738, 59

\bibitem[{{Shen} \& {Turner}(2008)}]{Shen2008}
{Shen}, Y., \& {Turner}, E.~L. 2008, \apj, 685, 553

\bibitem[{{Skrutskie} {et~al.}(2006){Skrutskie}, {Cutri}, {Stiening},
  {Weinberg}, {Schneider}, {Carpenter}, {Beichman}, {Capps}, {Chester},
  {Elias}, {Huchra}, {Liebert}, {Lonsdale}, {Monet}, {Price}, {Seitzer},
  {Jarrett}, {Kirkpatrick}, {Gizis}, {Howard}, {Evans}, {Fowler}, {Fullmer},
  {Hurt}, {Light}, {Kopan}, {Marsh}, {McCallon}, {Tam}, {Van Dyk}, \&
  {Wheelock}}]{Skrutskie2006}
{Skrutskie}, M.~F., {et~al.} 2006, \aj, 131, 1163

\bibitem[{{Smith} {et~al.}(2007){Smith}, {Collier Cameron}, {Christian},
  {Clarkson}, {Enoch}, {Evans}, {Haswell}, {Hellier}, {Horne}, {Irwin}, {Kane},
  {Lister}, {Norton}, {Parley}, {Pollacco}, {Ryans}, {Skillen}, {Street},
  {Triaud}, {West}, {Wheatley}, \& {Wilsons}}]{Smith2007}
{Smith}, A.~M.~S., {et~al.} 2007, in ASP Conference Series, Vol. 366,
  Transiting Extrapolar Planets Workshop, ed. C.~{Afonso}, D.~{Weldrake}, \&
  T.~{Henning}, 152

\bibitem[{{Southworth} {et~al.}(2009){Southworth}, {Hinse}, {J{\o}rgensen},
  {Dominik}, {Ricci}, {Burgdorf}, {Hornstrup}, {Wheatley}, {Anguita}, {Bozza},
  {Novati}, {Harps{\o}e}, {Kj{\ae}rgaard}, {Liebig}, {Mancini}, {Masi},
  {Mathiasen}, {Rahvar}, {Scarpetta}, {Snodgrass}, {Surdej}, {Th{\"o}ne}, \&
  {Zub}}]{Southworth2009}
{Southworth}, J., {et~al.} 2009, \mnras, 396, 1023

\bibitem[{{Tamuz} {et~al.}(2005){Tamuz}, {Mazeh}, \& {Zucker}}]{Tamuz2005}
{Tamuz}, O., {Mazeh}, T., \& {Zucker}, S. 2005, \mnras, 356, 1466

\bibitem[{{Wheatley} {et~al.}(2013){Wheatley}, {Pollacco}, {Queloz}, {Rauer},
  {Watson}, {West}, {Chazelas}, {Louden}, {Walker}, {Bannister}, {Bento},
  {Burleigh}, {Cabrera}, {Eigm{\"u}ller}, {Erikson}, {Genolet}, {Goad},
  {Grange}, {Jord{\'a}n}, {Lawrie}, {McCormac}, \& {Neveu}}]{Wheatley2013}
{Wheatley}, P.~J., {et~al.} 2013, in European Physical Journal Web of
  Conferences, Vol.~47, Hot Planets and Cool Stars, 13002

\bibitem[{Wright {et~al.}(2012)Wright, Marcy, Howard, Johnson, Morton, \&
  Fischer}]{Wright2012}
Wright, J., Marcy, G., Howard, A., Johnson, J.~A., Morton, T., \& Fischer,
  D.~A. 2012, \apj, 753, 160

\bibitem[{{Yanagisawa} {et~al.}(2006){Yanagisawa}, {Shimizu}, {Okita},
  {Nagayama}, {Sato}, {Koyano}, {Okada}, {Iwata}, {Uraguchi}, {Watanabe},
  {Yoshida}, {Okumura}, {Nakaya}, \& {Yamamuro}}]{Yanagisawa2006}
{Yanagisawa}, K., {et~al.} 2006, in Society of Photo-Optical Instrumentation
  Engineers (SPIE) Conference Series, Vol. 6269, Society of Photo-Optical
  Instrumentation Engineers (SPIE) Conference Series

\bibitem[{{Young} {et~al.}(1989){Young}, {Skumanich}, {Stauffer}, {Harlan}, \&
  {Bopp}}]{Young1989}
{Young}, A., {Skumanich}, A., {Stauffer}, J.~R., {Harlan}, E., \& {Bopp}, B.~W.
  1989, \apj, 344, 427

\end{thebibliography}
